\documentclass[aps,prb,twocolumn,showpacs,groupedaddress]{revtex4}
\usepackage{amssymb}

\input{tcilatex}

\begin{document}

\title{Thermodynamic Properties and Pressure Effect on the Superconductivity
in CaAlSi and SrAlSi}
\author{B. Lorenz$^{1}$, J. Cmaidalka$^{1}$, R. L. Meng$^{1}$, and C. W. Chu$%
^{1,2,3}$}
\affiliation{$^{1}$Department of Physics and TCSAM, University of Houston, Houston, TX
77204-5932}
\affiliation{$^{2}$Lawrence Berkeley National Laboratory, 1 Cyclotron Road, Berkeley, CA
94720}
\affiliation{$^{3}$Hong Kong University of Science and Technology, Hong Kong, China}
\date{\today }

\begin{abstract}
Superconductivity in the C32 compounds CaAlSi and SrAlSi is investigated by
heat capacity measurements as well as high-pressure ac-susceptibility
experiments. The heat capacity in the superconducting state is well
explained by the BCS theory for weak (SrAlSi) and enhanced coupling strength
(CaAlSi). The magnitudes of the superconducting gaps are estimated as >2.7
meV (CaAlSi) and >1.4 meV (SrAlSi). The pressure effect on the
superconducting transition temperature, $dlnT_{c}/dp$, is positive in CaAlSi
(+0.026 $GPa^{-1}$) but negative in SrAlSi (-0.024 $GPa^{-1}$). The opposite
sign of the pressure coefficients of $T_{c}$ indicates distinct differences
in the electronic structure and the density of states at the Fermi energy of
the two compounds in agreement with recent band structure calculations.
\end{abstract}

\maketitle











\section{Introduction}

Since the discovery of superconductivity at 39 K in $MgB_{2}$, the attention
has turned to other intermetallic compounds with similar structure or
lattice symmetry. Besides $MgB_{2}$ only a few binary compounds between the
IIA and IIIB elements crystallize in the hexagonal $C32$ structure, i.e. $%
CaGa_{2}$, $SrGa_{2}$, and $BaGa_{2}$. These gallium-based binary compounds
are not superconducting.\cite{1} However, a new class of pseudo-ternary
compounds with $C32$ structure, $AX_{2-x}Si_{x}$ with $A=Ca,Sr,Ba$ and $%
X=Ga,Al$, was recently synthesized and characterized.\cite{1,2,3,4,5,6} The
pseudo-ternary compounds are derived from the binary ones by partially
substituting $X$ (e.g. $Ga$) by silicon which results in a change of the
lattice parameters\cite{1,4} and, presumably, of the electronic structure.
The compounds are isostructural to $MgB_{2}$ where the $A$-ions occupy the $%
Mg$ sites and the $X$/$Si$ ions are randomly distributed among the boron
sites in the honeycomb planes of the $C32$ structure. The search for
superconductivity was successful for most of the materials mentioned above,
however, stoichiometric $BaAlSi$ is not superconducting above 2 K (the
measuring limit of previous investigations). Magnetization measurements show
the characteristic features of type II superconductors.\cite{2} The $C32$
structure is stable within a wide range of $Al$:$Si$ and $Ga$:$Si$ ratios
and superconductivity was observed in a range from 0.6 to 2.3 ($Al$:$Si$)
and 0.3 to 1.9 ($Ga$:$Si$), respectively.\cite{1,4} The maximum $T_{c}$'s
were found near the stoichiometric 1:1:1 composition ($x$=1) in $%
CaAl_{2-x}Si_{x}$, $SrGa_{2-x}Si_{x}$, and $BaGa_{2-x}Si_{x}$ but at $x$%
=0.85 in $CaGa_{2-x}Si_{x}$.

Because of the structural similarity to the famous sister compound, $MgB_{2}$%
, the question has been raised whether the properties and mechanisms leading
to superconductivity in the pseudo-ternary compounds are of the same origin
as those in magnesium diboride and why their $T_{c}$ is so much lower (the
highest $T_{c}$ $\thickapprox $ 7.9 K was found in $CaAlSi$). We have shown,
based on thermoelectric power measurements, that the majority of charge
carriers in the pseudo-ternary compounds are electron like in contrast to
the $\sigma $-hole carriers in $MgB_{2}$.\cite{1,4} This conclusion is
supported by recent band structure calculations showing that the
electron-like d-states of the alkaline-earth metals contribute up to 60 \%
to the total density of states at the Fermi energy.\cite{7}

It is interesting to note that there is no obvious correlation of $T_{c}$
with the mass of the alkaline-earth ion in the compound $A$-$GaSi$. However,
in $A$-$AlSi$ $T_{c}$ decreases systematically with increasing mass of the $%
A $-ion from 7.9 K ($A=Ca$) to 5 K ($A=Sr$) and $BaAlSi$ is not
superconducting above 2 K.\cite{4,6} Imai et al.\cite{6} attributed the
trend of $T_{c}$ in the $A$-$AlSi$ system to a decrease of the density of
states, $N(E_{F})$, due to the increasing mass of the $A$-ion assuming
thereby that the Debye temperature (the pre-factor in the BCS formula for $%
T_{c}$) does not change if $Ca$ is replaced by $Sr$ and $Ba$. This
assumption, however, is not justified because, unlike in $MgB_{2}$, the
electron-phonon interaction leading to the formation of Cooper pairs and
superconductivity is expected to involve the vibrations of the
alkaline-earth ions and their mass should affect the superconducting $T_{c}$%
. The latter argument is supported by the results of recent band structure
calculations for $A$-$AlSi$ indicating that $N(E_{F})$ actually increases if 
$Ca$ is replaced by $Sr$ or $Ba$.\cite{7} Furthermore, the alkaline-earth
d-states contribute to the Fermi surface to a large extent and there is no
reason to assume that they do not participate in the electron-phonon
coupling. Additional investigations are needed to reveal the physical nature
of the superconducting state.

We, therefore, have conducted measurements of the specific heat as well as
the effect of hydrostatic pressure on the superconducting transitions of $%
CaAlSi$ and $SrAlSi$. The heat capacity data are useful to (i) measure the
thermodynamic properties and confirm the bulk nature of superconductivity,
(ii) compare the specific heat in the normal and superconducting states with
the standard BCS theory,\cite{13,8} and (iii) extract microscopic
parameters, e.g. the superconducting gap. The application of pressure is of
interest because an extraordinary large superconducting $T_{c}$ of 14 K was
reported for a high pressure phase of the binary silicide $CaSi_{2}$.\cite{9}

The heat capacity data clearly show the existence of bulk superconductivity
in $CaAlSi$ and $SrAlSi$, consistent with previous magnetic measurements.%
\cite{1,2} The temperature dependence of the specific heat of $SrAlSi$ fits
well the BCS theory in the weak-coupling limit. Deviations from the BCS
model in the case of $CaAlSi$ indicate a slightly stronger coupling. With
the application of hydrostatic pressure $T_{c}$ increases in $CaAlSi$ but it
decreases in $SrAlSi$. The opposite signs of the pressure coefficients of $%
T_{c}$ are explained in terms of the distinct difference between the
electronic structures of both compounds.

\section{Experimental}

The pseudo-ternary intermetallic compounds, $CaAlSi$ and $SrAlSi$, have been
synthesized by argon arc melting with appropriate amounts of $Ca$ (99 \%), $%
Sr$ (99 \%), $Al$ (99.99 \%), and $Si$ (99.99 \%) as described earlier.\cite%
{1} The x-ray powder diffraction spectra show the $C32$ hexagonal structure
with lattice constants (a,c) = (4.189, 4.400 \AA ) for $CaAlSi$ and (a,c) =
(4.220, 4.754 \AA ) for $SrAlSi$. Magnetization and transport measurements
indicate the onset of superconductivity at 7.9 K and 5.0 K in $CaAlSi$ and $%
SrAlSi$, respectively.\cite{1,4}

The heat capacity was measured for both samples (typical mass of 25 mg)
between room temperature and 2.2 K employing the Physical Property
Measurement System (Quantum Design). The relative error of the measurement
(i.e. the statistical error of the fitting procedure) was of the order of
0.1 to 0.2 \%. The absolute accuracy of the measurement was tested in
particular at low temperatures by comparing the measured heat capacity of
high purity (99.9999 \%) lead with the available data.\cite{10} The
agreement of the lead measurement with the literature data was within the
statistical data fluctuation.

The pressure effect on the superconducting transitions of $CaAlSi$ and $%
SrAlSi$ was investigated by ac susceptibility measurements at pressures up
to 2 GPa. A dual coil system was mounted directly to the sample. The ac
susceptibility was measured using the mutual inductance bridge (LR700,
Linear Research). Pressure was generated in a beryllium-copper
piston-cylinder clamp.\cite{11} The sample was mounted in a Teflon container
filled with a 1:1 mixture of Fluorinert FC70 and FC77 as a hydrostatic
pressure transmitting medium. The pressure was measured in situ at 7 K by
monitoring the shift of the superconducting $T_{c}$ of a high purity
(99.9999 \%) lead manometer. The temperature above 45 K was measured by a
thermocouple inside the Teflon container and, at low temperatures, by a
germanium resistor built into the pressure cell close to the sample
position. Data have been taken upon loading and unloading cycles.

\section{Results and Discussion}

\subsection{Heat capacity measurements}

Figs. 1a and 1b show the low temperature heat capacity data, $C_{p}/T$ vs. $%
T,$ for $CaAlSi$ and $SrAlSi$, respectively. At zero magnetic field (open
circles) $C_{p}$ shows a sharp peak at the superconducting transition as
expected for bulk superconductors with a narrow transition width of 0.4 K ($%
CaAlSi$) and 0.2 K ($SrAlSi$). To compare with the normal state heat
capacity the superconductivity was suppressed by a 7 Tesla magnetic field.
The corresponding data are shown by open squares in Fig. 1. The high field
data can be used to estimate the superconducting component of $C_{p}$ by
subtracting the lattice contribution from the zero field $C_{p}$ below $%
T_{c} $. In general $C_{p}$ is given by the sum of the electronic ($C_{e}$)
and the lattice ($C_{l}$) contributions,

\begin{equation}
C_{p}(T)=C_{e}(T)+C_{l}(T)  \tag{1}
\end{equation}

with $C_{e}(T)=C_{e}^{(n)}(T)=\gamma T$ \ in the normal state and $%
C_{e}(T)=C_{e}^{(s)}(T)$ in the superconducting state. The values of $%
C_{e}^{(s)}(T)$ in the BCS theory for weak coupling were numerically
calculated and tabulated by M\"{u}hlschlegel.\cite{13} The characteristic
coefficient $\gamma $ of $C_{e}^{(n)}$ is proportional to the density of
states (DOS) at the Fermi energy, $N(E_{F})$. The lattice heat capacity at
low temperatures can be expanded in a power series of $T$ (Debye model)
starting with the third order term, $T^{3}$. Normal state electronic and
lattice contributions are commonly separated in the $C_{p}/T$ vs. $T^{2}$
plot where the coefficient $\gamma $ is obtained as the linear extrapolation
of $C_{p}/T$ to zero temperature and the slope at low temperatures is used
to calculate the Debye temperature, $\Theta _{D}$. Using the high field data
of $C_{p}$ we estimate $\gamma =5.04\ mJ/molK,\ \Theta _{D}=226\ K$ and $%
\gamma =5.48\ mJ/molK,\ \Theta _{D}=218\ K$ for $CaAlSi$ and $SrAlSi$,
respectively. It is interesting to note that $\gamma $ of $SrAlSi$ is indeed
larger than the value for $CaAlSi$ indicating that $N(E_{F})$ increases with
the substitution of $Ca$ by $Sr$ in qualitative agreement with the band
structure calculations.\cite{7} However, the increase is moderate and far
smaller than the factor of two predicted by the theory.

The electronic heat capacity in the superconducting state, $C_{e}^{(s)}(T)$,
is compared with the BCS data (dotted line) in the insets of Fig. 1. The
thermodynamic $T_{c}$ was estimated by an entropy conserving construction as 
$T_{c}=7.7\ K$ ($CaAlSi$) and $T_{c}=4.9\ K$ ($SrAlSi$). These values are
slightly lower than the values determined from resistivity measurements or
from the onset of the diamagnetic drop of the magnetic susceptibility.
However, this is to be expected since the thermodynamic $T_{c}$ represents
an averaged bulk value whereas resistivity as well as the onset of
diamagnetism are both sensitive to the first percolating path of
superconducting volume in the sample. The normalized electronic heat
capacity, $C_{e}^{(s)}(T)/C_{e}^{(n)}(T)$, \ is displayed as a function of $%
T/T_{c}$. There is an excellent agreement with the BCS data in the case of $%
SrAlSi$ (Fig. 1b), i.e. the BCS model in weak coupling limit describes the
superconducting state very well in the temperature range accessible in our
experiments ($T>2.2\ K$). There are deviations of $C_{e}^{(s)}(T)$ from the
BCS data in case of $CaAlSi$. The heat capacity falls below the BCS values
at low $T$ but is enhanced close to $T_{c}$ resulting in a larger relative
jump of $C_{p}$, $\delta C_{p}(T_{c})/\gamma T_{c}\thickapprox 2$ (BCS value
= 1.43). These deviations may be explained by a stronger electron phonon
coupling in $CaAlSi$ as compared to $SrAlSi$. The thermodynamic consistency
of the heat capacity data was checked by integrating the entropy difference, 
$(C_{e}^{(s)}(T)-C_{e}^{(n)}(T))/T$, from $T=0$ to $T=T_{c}$. For $CaAlSi$
this integral is close to zero as expected. For $SrAlSi$ the integration
cannot be carried out with sufficient accuracy because of the experimental
restrictions (the lowest accessible temperature, 2.2 K, is just little less
than 50 \% of $T_{c}$). The superconducting gap at zero temperature, $%
2\Delta $, may be estimated from the heat capacity at low temperatures by
plotting $lnC_{e}^{(s)}(T)$ vs. $1/T$ (Arrhenius plot). The low temperature
slopes estimated from this plot yield values of $2\Delta =1.4\ meV$ and $%
2\Delta =2.7\ meV$ for $SrAlSi$ and $CaAlSi$, respectively. However, because
of the limited temperature range of our data ($T>$2.2\ K) these values are
certainly underestimated and should be considered as a lower limit of the
gap parameter. A more accurate estimate requires heat capacity measurements
at far lower temperatures. The BCS value of the gap is related to $T_{c}$ by 
$2\Delta _{BCS}=3.53\ k_{B}\ T_{c}$. For $SrAlSi$ with $T_{c}=4.9\ K$ we get 
$2\Delta _{BCS}=1.49\ meV$ which is only slightly larger than our estimate.
Taking into account the almost perfect agreement of $C_{e}^{(s)}(T)$ with
the BCS function over the full experimental temperature range, the
superconducting gap value for $SrAlSi$ can be assumed to be close to the BCS
value of $1.49\ meV$. However, the calculated BCS gap parameter for $CaAlSi$%
, $2\Delta _{BCS}=2.34\ meV$, is clearly smaller than the value of $2.7\ meV$
estimated as a lower limit of $2\Delta $ from the Arrhenius plot of $%
C_{e}^{(s)}(T)$. This enhancement of the gap with respect to the BCS
parameter is a further indication of an increased electron-phonon coupling.
A summary of all relevant parameters and the comparison to the BCS weak
coupling values is given in Table 1.

\begin{table}[tbp]
\caption{The characteristic normal and superconducting state parameters of
CaAlSi and SrAlSi as compared to the weak coupling BCS theory.}
\label{1}%
\begin{ruledtabular}
\begin{tabular}{llll}
& CaAlSi & SrAlSi & BCS-Theory \\ 
$\gamma \ [mJ/molK^{2}]$ & 5.04 & 5.42 & - \\ 
$\theta_D \ [K]$ & 226 & 218 & - \\
$T_{c}\ [K]$ & 7.7 & 4.9 & - \\ 
$2\Delta \ [meV]$ & 2.7 & 1.49 & - \\ 
$\delta C_{p}(T_{c})/\gamma T_{c}$ & 2.0 & 1.4 & 1.43 \\ 
$2\Delta /k_{B}T_{c}$ & 4.07 & $\sim $3.5 & 3.53%
\end{tabular}
\end{ruledtabular}
\end{table}

The temperature dependence of the heat capacity in the superconducting state
provides strong evidence that the superconductivity found in $SrAlSi$ and $%
CaAlSi$ is well explained by the standard BCS theory for weak and enhanced
coupling, respectively. Furthermore, the data confirm the bulk nature of the
superconductivity in these pseudo-ternary $C32$ compounds.

It appears interesting to compare the present results with the
superconductivity found at relatively high temperatures in binary compounds
with similar structure. Superconductivity at 14 K was recently detected in $%
CaSi_{2}$ at pressures above 14 GPa.\cite{9} The (tetragonal) structure of
this high pressure phase is very similar to the $AlB_{2}$ structure but the
honeycomb planes of the $Si$ ions show still a nonvanishing buckling even at
the highest pressures.\cite{14} No thermodynamic data such as heat capacity
are known for this superconducting high pressure phase. In contrast,
extensive research has been conducted to investigate the thermodynamic
properties in the superconducting state of $MgB_{2}$. The heat capacity of $%
MgB_{2}$ shows an abnormal temperature dependence below $T_{c}$.\cite{15}
Whereas the value of $C_{e}^{(s)}(T)$ at and right below $T_{c}$ is smaller
than the corresponding BCS function it raises above the BCS data below about
20 K followed by an exponential drop at lower temperature. From this unusual
behavior the existence of two superconducting gaps in $MgB_{2}$ was
suggested and later confirmed by many alternative experiments. The present
data for $CaAlSi$ and $SrAlSi$ show no evidence for a similar scenario
within the experimental temperature range of $T>$2.2 K, i.e. $T>0.44\ T_{c}$
for $SrAlSi$ and $T>0.27\ T_{c}$ for $CaAlSi$. The existence of an anomaly
at far lower temperature, similar to $MgB_{2}$, cannot be completely
excluded from the current investigation. However, the almost perfect
agreement of $C_{e}^{(s)}(T)$ with the BCS data in the case of $SrAlSi$ as
well as the enhancement of $C_{e}^{(s)}(T)$ close to $T_{c}$ of $CaAlSi$
(instead of the depletion observed in $MgB_{2}$) suggest a less complex
structure of the superconducting gap as compared to magnesium diboride.

\subsection{Pressure effect on the superconducting transitions in CaAlSi and
SrAlSi}

The effect of hydrostatic pressure on the superconducting $T_{c}$ is of
special interest since it can help to reveal the intrinsic mechanisms of
superconductivity. In $MgB_{2}$, for example, it could be shown that the
negative pressure coefficient of $T_{c}$ was most compatible with the
strong-coupling model (McMillen) of phonon mediated superconductivity so
that alternative explanations appeared to be less favorable.\cite{12}

The pressure effect on $T_{c}$ of $CaAlSi$ and $SrAlSi$ was investigated by
measuring the real part of the ac susceptibility, $\chi _{ac}^{\prime }$.
The data in Fig. 2 show the shift of the diamagnetic signal of $\chi
_{ac}^{\prime }$ with increasing pressure. Surprisingly, the pressure effect
on $CaAlSi$ and $SrAlSi$ is opposite in sign. The superconducting transition
temperature (measured at the onset of the diamagnetic signal) of $CaAlSi$
increases with pressure whereas $T_{c}$ of $SrAlSi$ clearly decreases. The $%
T_{c}$'s as a function of pressure for both compounds are shown in Fig. 3.
The increase of $T_{c}$ in $CaAlSi$ (Fig. 3a) is slightly nonlinear with an
initial slope of $0.21\ K/GPa$. In contrast, Fig. 3b shows an almost perfect
linear decrease of $T_{c}$ for $SrAlSi$ with a coefficient of $-0.12\ K/GPa$%
. Although the magnitude of the relative pressure coefficient, $%
dln(T_{c})/dp $, is almost the same for both compounds, the opposite sign
deserves further consideration.

For a qualitative discussion we use the BCS equation for $T_{c}$\cite{8} 
\begin{equation}
k_{B}T_{c}=1.13\ \text{%
h{\hskip-.2em}\llap{\protect\rule[1.1ex]{.325em}{.1ex}}{\hskip.2em}%
}\omega _{D}\exp \left( -1/\lambda \right)  \tag{2}
\end{equation}

$\ $%
h{\hskip-.2em}\llap{\protect\rule[1.1ex]{.325em}{.1ex}}{\hskip.2em}%
\ $\omega _{D}$ is a measure of the characteristic energy scale for the
phonons and $\lambda =V_{0}N(E_{F})$. $V_{0}$ is the matrix element of the
effective interaction. Assuming that the pressure dependence of $V_{0}$ is
negligible, the pressure coefficient of $T_{c}$ is derived as

\begin{equation}
\frac{d\ln T_{c}}{dp}=\frac{d\ln \omega _{D}}{dp}+\frac{1}{V_{0}N(E_{F})}\ 
\frac{d\ln \left[ V_{0}N(E_{F})\right] }{dp}  \tag{3}
\end{equation}

There are obviously two contributions to the pressure coefficient of $T_{c}$%
: one from the sole phonon system, $\frac{d\ln \omega _{D}}{dp}$, and the
second one involving the electronic system. In the weak-coupling limit of
the BCS model both terms contribute to the total pressure coefficient. The
effect of pressure on the phonon system usually results in an increase of
the average phonon energy, i.e. in a ``hardening'' of phonon modes. Only in
rare cases, close to a structural transition, phonon ``softening'' may occur
but usually only for some particular modes. Therefore, the first term in
equation (3) yields a positive contribution to the total pressure
coefficient. The pressure effect on the density of states, however, is more
complex. The compression of the lattice by pressure in general causes an
increase of the bandwidth which in turn will result in an average decrease
of $N(E)$ since the total number of states in a band is fixed. In addition, $%
V_{0}$\ tends to decrease with pressure since it is inversely proportional
to the average square of the phonon frequency. This negative contribution to 
$\frac{d\ln T_{c}}{dp}$ will compete with the phonon hardening effect.
However, the application of pressure may also cause slight changes in the
band structure and/or a shift of the Fermi energy, $E_{F}$, and a change of $%
N(E_{F})$. Therefore, within the BCS theory the pressure coefficient of $%
T_{c}$ can be of either sign, depending on which of the contributions to
equation (3), positive or negative, dominate.

In both compounds, $CaAlSi$ and $SrAlSi$, the phonon contribution to the
pressure coefficient is expected to be positive. The magnitude of $\frac{%
d\ln \omega _{D}}{dp}$\ is not known so far but it can be estimated from
Raman or infrared spectroscopy at high pressures. The electronic term in
equation (3) needs a careful consideration. Recent band structure
calculations have shown striking differences in $N(E)$ in $CaAlSi$ and $%
SrAlSi$.\cite{7} In $CaAlSi$ the Fermi energy lies in a region where $N(E)$
is flat and relatively insensitive to small variations of $E_{F}$. This
results in a small electronic contribution to the total pressure coefficient
of $T_{c}$. In $CaAlSi$ the positive phonon term obviously outweighs the
electronic term which results in the observed positive $\frac{dT_{c}}{dp}$
as shown in Fig. 3a. The situation is different, however, in $SrAlSi$.
According to the calculations\cite{7} the density of states forms a very
narrow peak close to the Fermi energy. This makes $N(E_{F})$ extremely
susceptible to external perturbations. Any small change induced by pressure
(band broadening, shift of $E_{F}$) may result in a stronger decrease of $%
N(E_{F})$. The negative electronic contribution to the pressure coefficient
of $T_{c}$ therefore can dominate over the lattice term and the total $\frac{%
dT_{c}}{dp}$ becomes negative as observed in our experiments (Fig. 3b).

Although the above discussion is qualitative in nature it indicates that the
opposite pressure coefficients of $T_{c}$ in $CaAlSi$ and $SrAlSi$ uniquely
reflect the peculiarities of their electronic structure. They provide an
indirect proof of the distinct differences in the density of states at $%
E_{F} $ in both compounds, $CaAlSi$ and $SrAlSi$, as suggested by band
structure calculations. Furthermore, the positive pressure coefficient
observed in $CaAlSi$ cannot be explained by a sole change of the density of
states induced by pressure but the contribution due to phonon hardening has
to be taken into account. For a more quantitative comparison additional
experimental as well as theoretical work is needed. The measurement of
phonon frequencies as a function of pressure should provide a better
estimate of the phonon effect on $\frac{d\ln T_{c}}{dp}$. Improved band
structure calculations involving the effects of lattice compression and
including the phonon system (see for example ref. 16) may help to get a more
adequate understanding of the superconductivity in the pseudo-binary C32
intermetallic compounds.

\section{Summary and Conclusions}

We have investigated the superconductivity in the $C32$ intermetallic
compounds $CaAlSi$ and $SrAlSi$ by heat capacity and high-pressure ac
susceptibility measurements. Several parameters characterizing the normal
and superconducting states are extracted from the heat capacity. The
coefficients of the normal state electronic specific heat are estimated as $%
\gamma =5.04\ mJ/molK^{2}$ and $\gamma =5.42\ mJ/molK^{2}$ for $CaAlSi$ and $%
SrAlSi$, respectively. The values of the superconducting gap are calculated
from the electronic heat capacity in the superconducting state as $2.7\ meV$
($CaAlSi$) and $1.4\ meV$ ($SrAlSi$). The parameters are consistent with the
predictions of the BCS theory in the weak-coupling limit for $SrAlSi$ and
for enhanced coupling in the case of $CaAlSi$. The bulk nature of
superconductivity in $CaAlSi$ and $SrAlSi$ is confirmed.

The effect of hydrostatic pressure on $T_{c}$ is positive in the case of $%
CaAlSi$ ($\frac{d\ln T_{c}}{dp}=0.026\ GPa^{-1}$) but negative for $SrAlSi$ (%
$\frac{d\ln T_{c}}{dp}=-0.024\ GPa^{-1}$). This opposite tendency of the
pressure shift of $T_{c}$ reflects the distinct differences in the density
of states near the Fermi energy in both compounds. The high pressure data
also underline the importance of the phonon frequency (or ionic mass) for
understanding the superconductivity in the system $AAlSi$ and, in
particular, the positive pressure effect observed in $CaAlSi$. With the
increasing mass of the alkaline-earth ion $T_{c}$ is systematically
suppressed, from 7.7\ K ($Ca$) to 4.9\ K ($Sr$) and to non-superconducting
above 2\ K ($Ba$). This tendency cannot be explained as a sole
density-of--states effect since our data show that $N(E_{F})$ increases when 
$Ca$ is replaced by $Sr$, in qualitative agreement with the band structure
calculations.

\begin{acknowledgments}
This work is supported in part by NSF Grant No. DMR-9804325, the T.L.L.
Temple Foundation, the John J. and Rebecca Moores Endowment, and the State
of Texas through the TCSAM at the University of Houston and at Lawrence
Berkeley Laboratory by the Director, Office of Energy Research, Office of
Basic Energy Sciences, Division of Materials Sciences of the U.S. Department
of Energy under Contract No. DE-AC03-76SF00098.
\end{acknowledgments}

\begin{figure}[tbp]
\caption{Low temperature heat capacity of (a) CaAlSi and (b) SrAlSi. The
open circles show $C_{p}$ measured in zero magnetic field. The normal state $%
C_{p}$ (open squares) was measured by suppressing superconductivity in a
magnetic field of 7 Tesla. The insets show the plot of the electronic
specific heat in the superconducting state, $C_{e}^{(s)}/C_{e}^{(n)}$ vs. $%
T/T_c$, with the BCS function (dotted line).}
\label{F1}
\end{figure}

\begin{figure}[tbp]
\caption{ac susceptibility of (a) CaAlSi and (b) SrAlSi measured at
different pressures. The data are normalized to the susceptibility value
right above $T_{c}$. The superconducting transition temperature was
estimated from the onset of the diamagnetic signal.}
\label{F2}
\end{figure}

\begin{figure}[tbp]
\caption{Pressure dependence of $T_{c}$ of (a) CaAlSi and (b) SrAlSi. The
filled squares and open circles are data taken at increasing and decreasing
pressure, respectively}
\label{F3}
\end{figure}


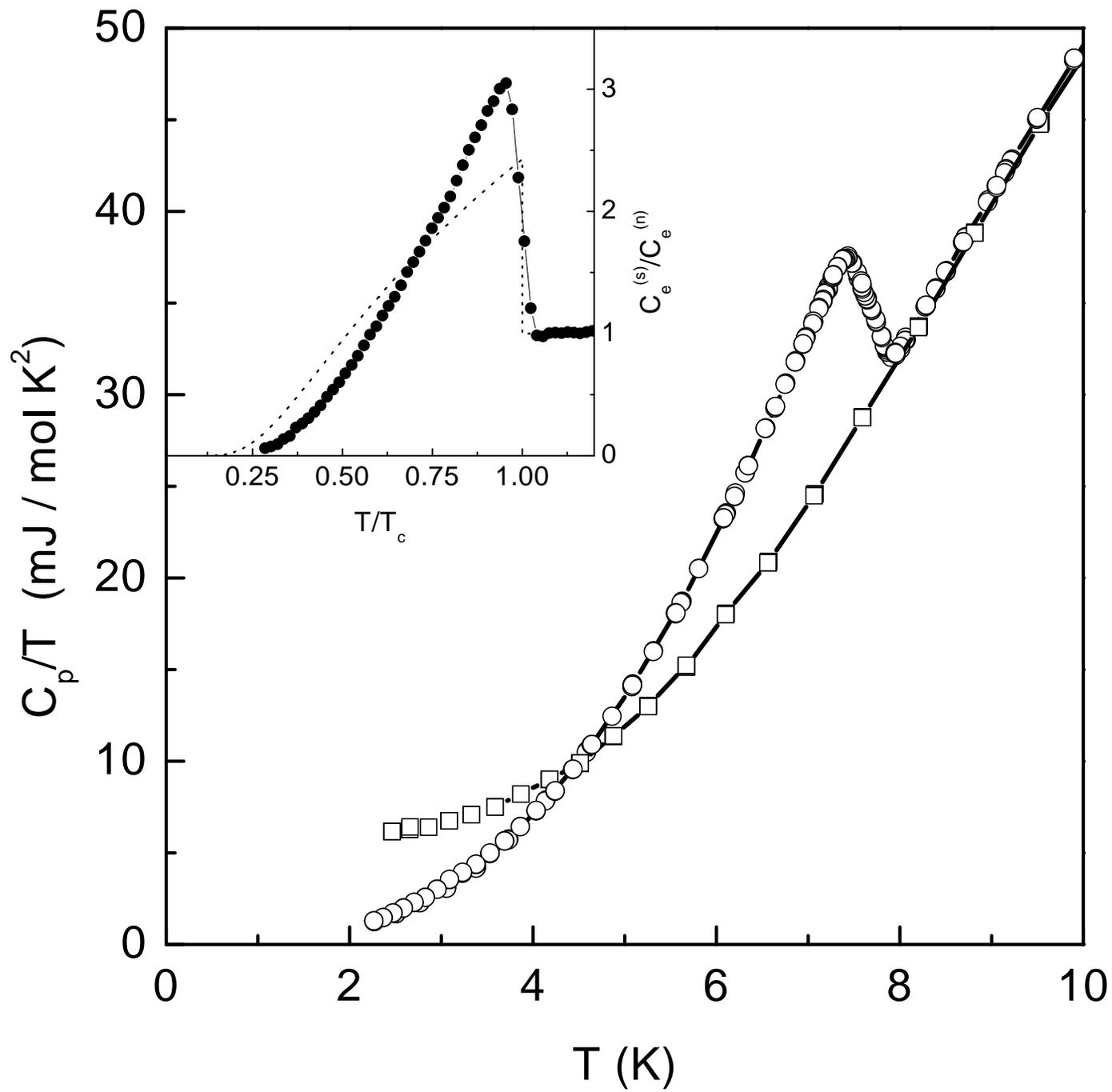
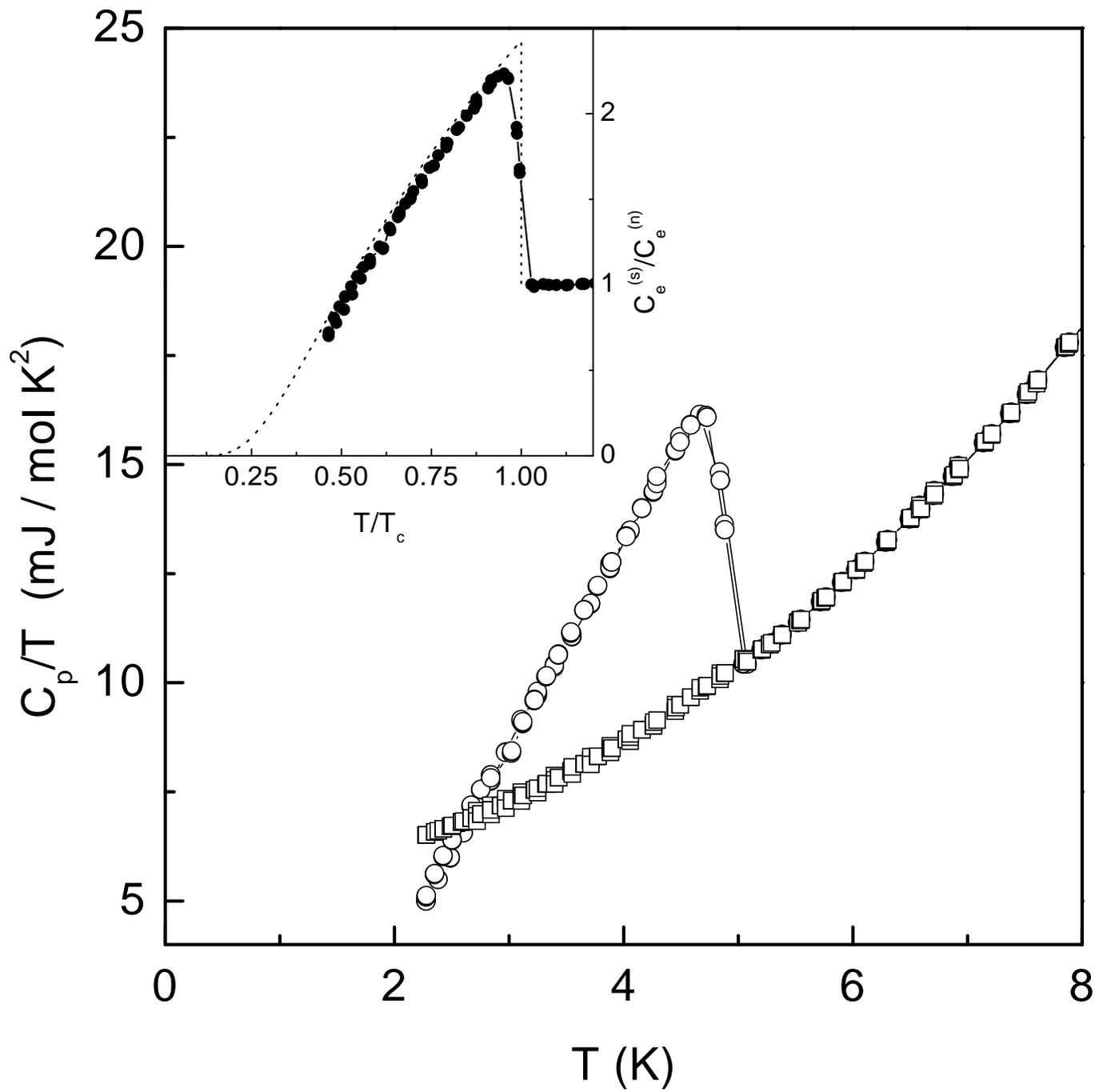
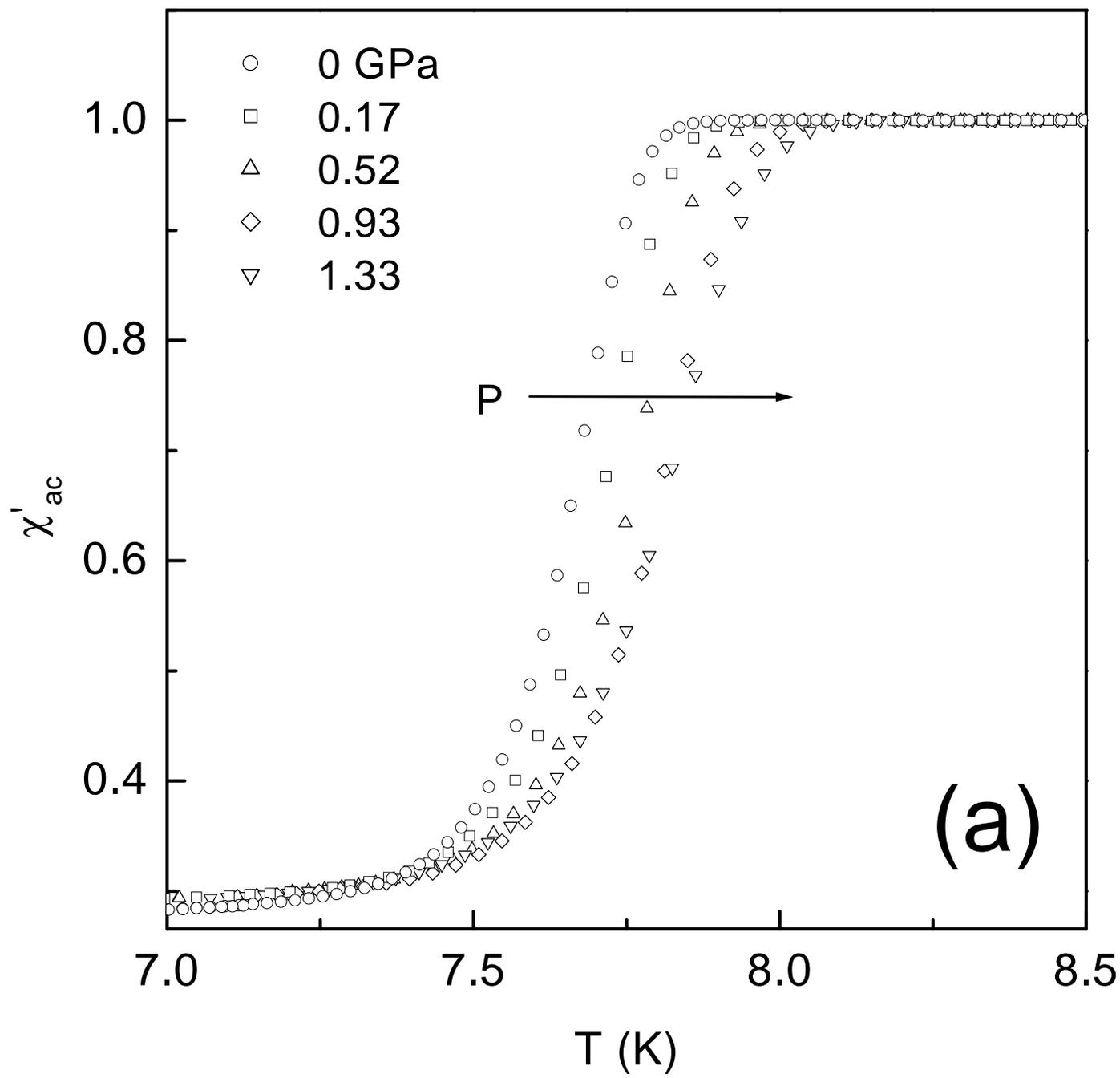
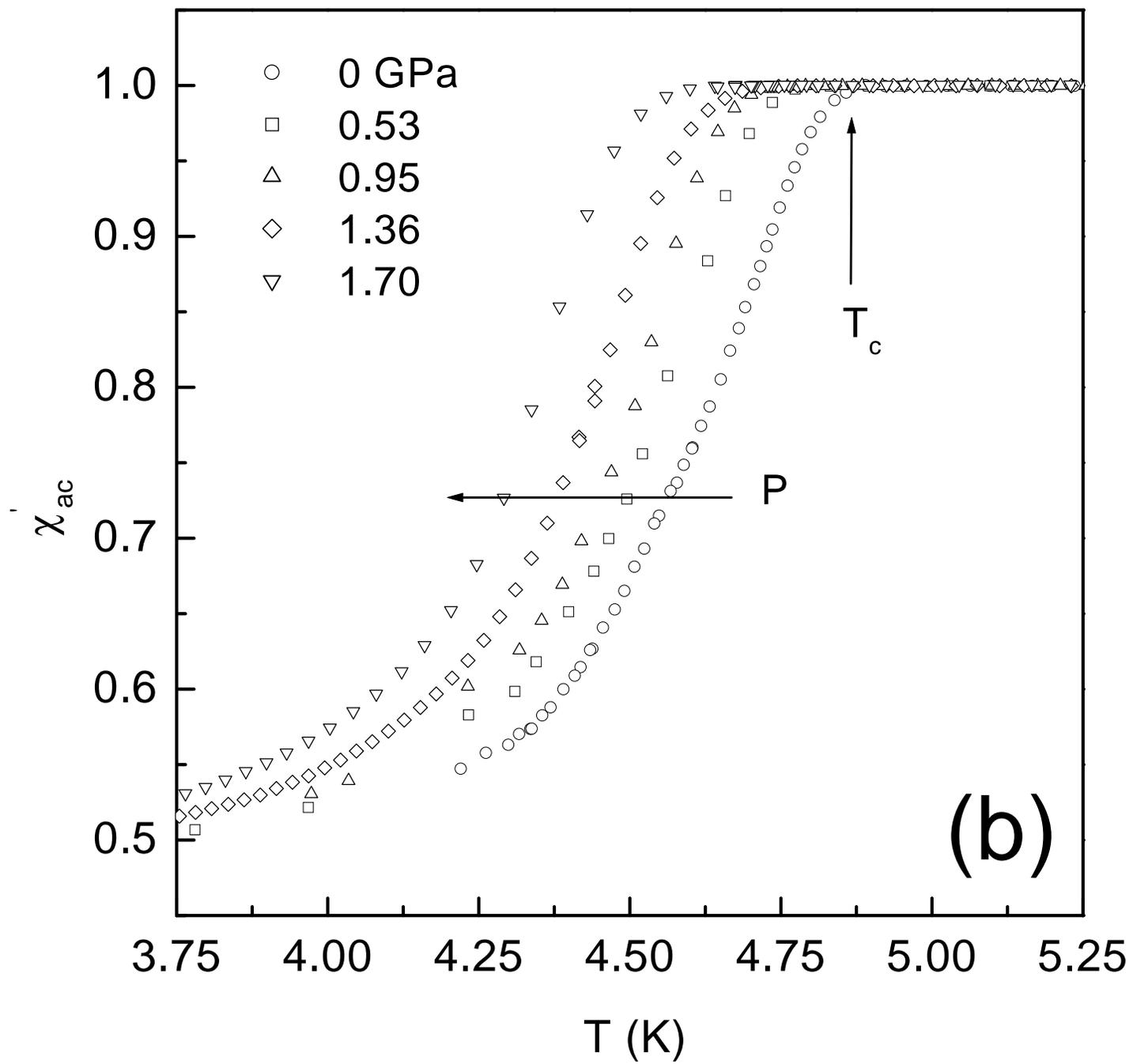
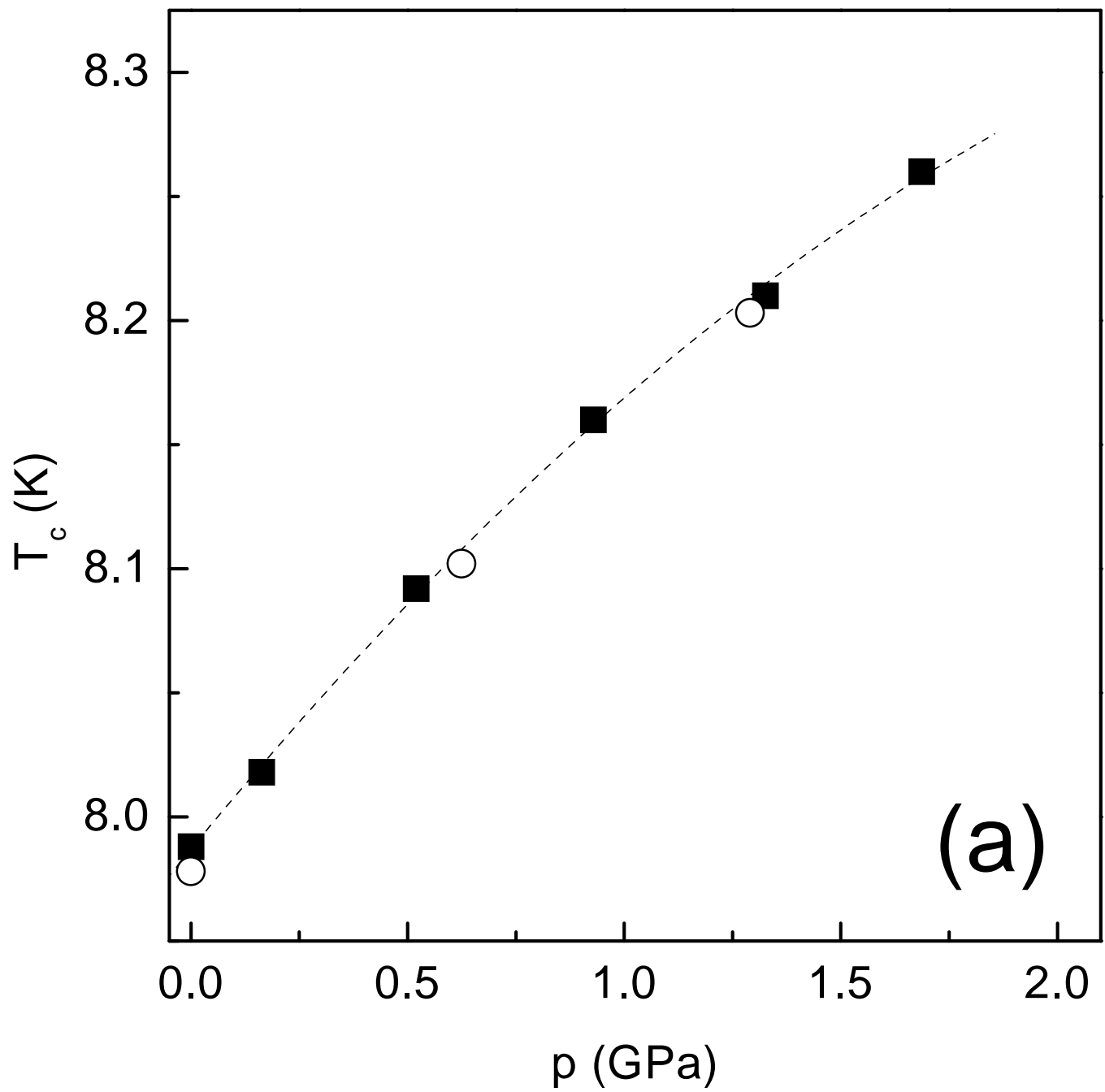
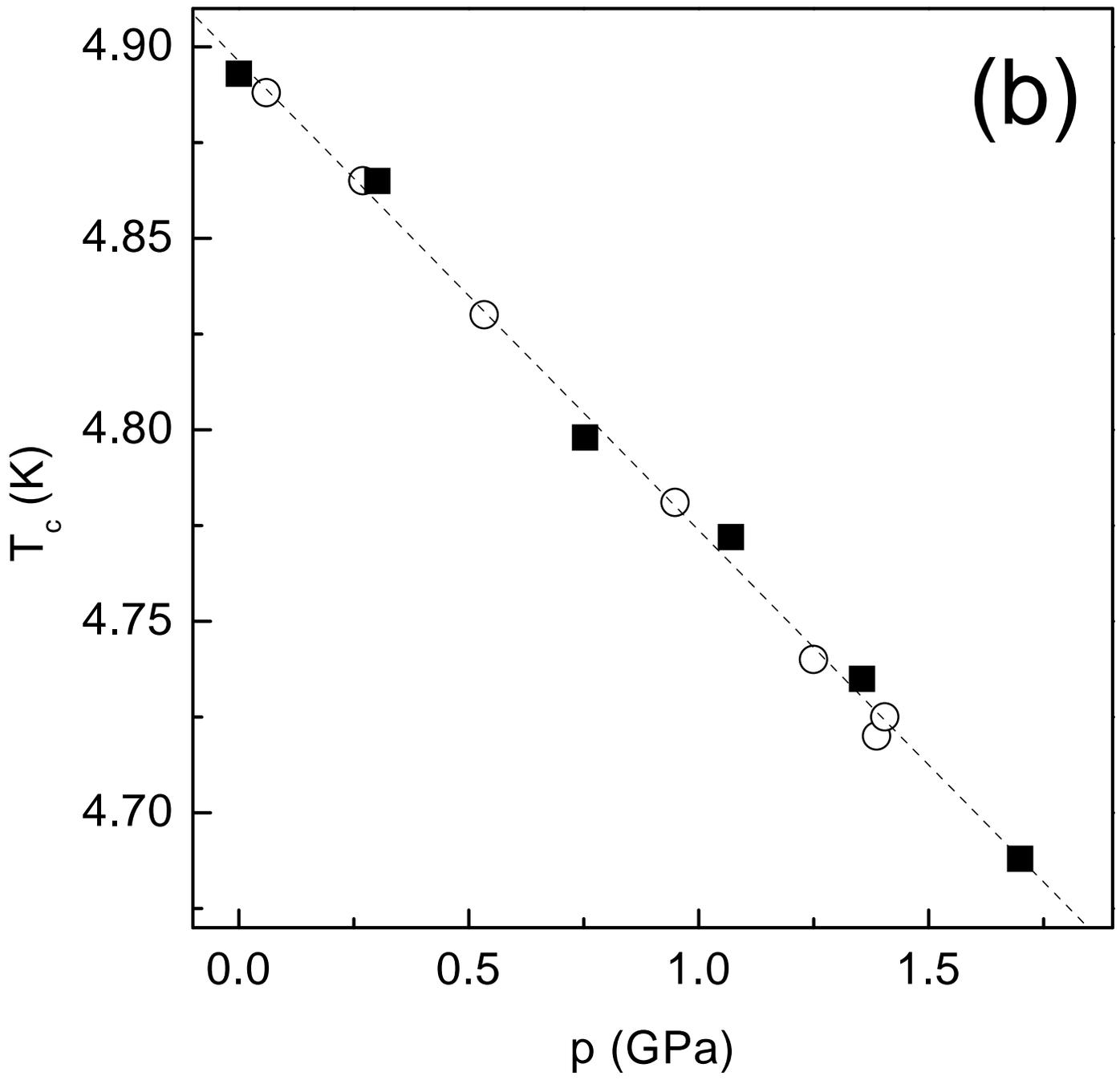
\end{document}